# Systematic validation of time-resolved diffuse optical simulators via non-contact SPAD-based measurements

Weijia Zhao[1], Linlin Li[1], Kaiqi Kuang[1], Yang Lin[2], Claudio Bruschini[2], Jiaming Cao[3,4], Ting Li[5], Edoardo Charbon[2,*] and Wuwei Ren[1,*]

[1] School of Information Science and Technology, ShanghaiTech University, Shanghai, China
[2] AQUA Laboratory, École Polytechnique Fédérale de Lausanne (EPFL), Rue de la Maladière 71, Neuchâtel, Switzerland
[3] Centre for Cognitive and Brain Sciences, Institute of Collaborative Innovation, University of Macau, Taipa, Macau SAR, China
[4] Department of Electrical and Computer Engineering, Faculty of Science and Technology, University of Macau, Taipa, Macau SAR, China
[5] Institute of Biomedical Engineering, Chinese Academy of Medical Sciences & Peking Union Medical College, Tianjin 300192, China

*Author to whom any correspondence should be addressed.

E-mail: renww@shanghaitech.edu.cn; edoardo.charbon@epfl.ch

## Abstract

**Objective**: Time-domain diffuse optical imaging (DOI) requires accurate forward models for photon propagation in scattering media. However, existing simulators lack comprehensive experimental validation, especially for non-contact configurations with oblique illumination. This study rigorously evaluates three widely used open-source simulators, including MMC (Mesh-based Monte Carlo), NIRFASTer, and Toast++ (the latter two are finite-element method (FEM)-based), using time-resolved experimental data. **Approach**: All simulations employed a unified mesh and point-source illumination. Virtual source correction was applied to FEM solvers for oblique incidence. A time-resolved DOI system with a 32 × 32 single-photon avalanche diode (SPAD) array acquired transmission-mode data from 16 standardized phantoms with varying absorption coefficient $\mu_a$ and reduced scattering coefficient $\mu_s'$. The simulation results were quantified across five metrics: spatial-domain (SD) precision, time-domain (TD) precision, oblique beam accuracy, computational speed, and mesh-density independence. **Results**: Among three simulators, MMC achieves superior accuracy in SD and TD metrics (SD MSE: 0.072 ± 0.053; TD MSE: 0.179 ± 0.080), and shows robustness across all optical properties. NIRFASTer and Toast++ demonstrate comparable overall performance. In general, MMC is optimal for accuracy-critical TD-DOI applications, while NIRFASTer and Toast++ suit scenarios prioritizing speed (e.g., image reconstruction) with sufficiently large $\mu_s'$. Besides, virtual source correction is essential for non-contact FEM modeling, which reduced average errors by > 34% in large-angle scenarios. **Significance**: This work provides benchmarked guidelines for simulator selection during the development phase of next-generation TD-DOI systems. Our work represents the first study to systematically validate TD simulators against SPAD array-based data under clinically relevant non-contact conditions, bridging a critical gap in biomedical optical simulation standards.

## 1. Introduction

Diffuse optical imaging (DOI) leverages non-invasive detection of transmitted diffusive photons and visualizes intrinsic optical properties or exogenous contrast agents within living tissues (Gibson et al., 2005), thereby enabling the retrieval of functional or molecular information of living animals or human beings, such as blood oxygenation (Jobsis, 1977) and targeted fluorescence distributions (Yi et al., 2014). Diverse DOI modalities have been proposed, including diffuse optical tomography (DOT) (Zhao et al., 2021), fluorescence molecular tomography (FMT) (Cai et al., 2020), spatial frequency domain imaging (SFDI) (Gioux et al., 2019) and diffuse correlation spectroscopy (DCS) (Carp et al., 2023), demonstrating a wide spectrum of applications covering disease diagnostics (Karthika et al., 2024), brain imaging (McMorrow et al., 2025), and preclinical small animal imaging (Schraven et al., 2024). According to the measurement configuration, DOI techniques can be classified into three modes: continuous-wave (CW), frequency-domain (FD), and time-domain (TD). Among these, TD-mode DOI offers the richest spatiotemporal information, potentially useful for improving spatiotemporal resolution, deepening detection penetration, and enhancing image contrast (Boas et al., 2016, Hillman, 2002, Kumar et al., 2008, Sutin et al., 2016). Recent technological advancements of time-resolved (TR) detection, such as single-photon avalanche diode (SPAD), have lowered the cost of the previously expensive TD DOI systems and improved the compactness and flexibility of those systems (Pifferi et al., 2016), unlocking new applications such as wearable devices for hemodynamic monitoring (Lacerenza et al., 2020).

Regardless of the difference between imaging modalities, the general goal of DOI is to recover certain image contrast, e.g., optical absorption or fluorescence, which requires knowledge of how light propagates in scattering tissues. The modelling process of predicting photon distributions within the medium is known as the forward model (Hoshi and Yamada, 2016). The forward model of DOI is based on either the radiative transfer equation (RTE) (Welch and Van Gemert, 2011) or diffusion equation (DE), a simplified form of RTE with diffusion approximation (Welch and Van Gemert, 2011). Despite the existence of analytical solutions for simplified imaging objects, RTE and DE are typically solved numerically, by using either Monte Carlo (MC) methods or finite element methods (FEM) (Yamada and Hasegawa, 1996). Accurate forward modelling is crucial for improving the imaging quality of DOI, particularly for the information-rich TD-mode systems such as TD-DOT (Bouza Domínguez and Bérubé-Lauzière, 2012, Bai et al., 2024) and TD-FMT (Chen and Intes, 2011, Gao et al., 2014). Therefore, experimental validation and comparative evaluation of TD diffuse optical simulators are of significant importance.

The validation of forward models for diffuse light propagation began in the 1980s. Flock et al. compared analytical diffusion theory against MC simulations. Results achieved from both simulators were further assessed via liquid phantoms measured by a CW-mode DOI system with optical fiber-based detection and photomultiplier tube (PMT) (Flock et al., 2002). Subsequently, Hillman et al. and Okada et al. conducted time-resolved measurements and benchmarked MC/FEM simulations against cylindrical phantom data acquired with fiber bundles connected to a PMT and a streak camera, respectively (Hillman et al., 2000, Okada et al., 1996). Jiang et al. assessed the simulation results obtained from MC and FEM using a CW-mode DOI system equipped with a CCD camera, where the experimental data were acquired using a silicone phantom in both reflection and transmission configurations (Jiang et al., 2020). Recently, simulation validations using time-resolved SPAD-based measurements were reported. Among these, Jiang et al. evaluated the time-of-flight (ToF) histogram acquired at the central pixel of a SPAD array with the simulation results from NIRFASTer, a FEM-based simulator (Jiang et al., 2021). Similarly, Kalyanov et al. used MC simulation to generate the ToF histograms, which were further validated using SPAD-based measurement at the central and peripheral pixels (Kalyanov et al., 2018). To date, most TD DOI studies suffer from incomplete validation, which relies on simulation-only verification or single-phantom tests, hindering accurate recovery of the targeted optical properties and downstream data analysis. Besides, previous validation work focuses on contact-mode configurations, where free-space illumination and detection were often neglected. Furthermore, there is no direct and experimental comparison among different TR diffuse light propagation simulators. To sum up briefly, a systematic validation of TR simulators based on non-contact experimental data with high temporal and spatial resolution is urgent for the development of next-generation TD DOI systems.

Our work bridges this gap by performing rigorous experimental validation on multiple routinely used TR simulators, which were all based on standardized homogeneous phantoms, SPAD array measurements, and non-contact detection considering various incident angles, enabling comprehensive multi-simulator benchmarking at SPAD array-level spatial resolution under a unified condition. We systematically evaluated two FEM tools (NIRFASTer (Dehghani), Toast++ (Schweiger and Arridge)) and an MC-based simulator (MMC (Fang)) with a homebuilt TD DOI system equipped with a SPAD array. The configurations for different simulators were consistent by using an identical 3D mesh and single-point illumination, which enables direct comparison between these simulators and real experiments. To improve the accuracy of FEM simulation, we introduced virtual source correction for an oblique incidence beam. All simulation results were quantified across five key dimensions (spatial-domain (SD) precision, TD precision, oblique beam accuracy, speed, mesh-density independence), providing an objective comparison under varied conditions.

## 2. Methods

### *2.1 Theory of diffuse light propagation modelling*

DOI requires an understanding of diffuse light propagation in highly scattering media such as biological tissues. The RTE is used as the gold standard for precisely predicting the radiative energy deposition in tissue by accounting for absorption, scattering, and anisotropy in the time domain, expressed as follows.

$$\frac{1}{c}\frac{\partial L(\vec{r},\hat{s},t)}{\partial t} = -\hat{s}\cdot\nabla L(\vec{r},\hat{s},t) - \mu_t(\vec{r})L(\vec{r},\hat{s},t) + \mu_s(\vec{r})\int_{4\pi} L(\vec{r},\hat{s}',t)P(\hat{s}'\cdot\hat{s})d\Omega' + S(\vec{r},\hat{s},t) \quad (1)$$

Where the change rate of the radiative energy $L(\vec{r},\hat{s},t)$ at position $\vec{r}$ and time $t$ in direction $\hat{s}$ is attributed to the comprehensive physical processes of photon propagation, absorption, scattering, and external sources. Here, $\mu_t(\vec{r}) = \mu_a(\vec{r}) + \mu_s(\vec{r})$ is the total interaction coefficient, where $\mu_a$ and $\mu_s$ represent the absorption and scattering coefficients, respectively. $P(\hat{s}'\cdot\hat{s})$ is the scattering phase function, and $S(\vec{r},\hat{s},t)$ represents the source term. Due to the high complexity of solving the RTE in an analytical way, a numerical solution based on MC methods is typically applied, which statistically simulates the transport behaviors of a large number of launched photons (Fang, 2010).

In highly scattering media, the RTE (Equation (1)) can be approximated by the DE, provided the photon transport distance exceeds one transport mean free path (TMFP) (Jacques and Pogue, 2008). FEM is employed to solve the DE by discretizing the domain into a finite number of simple geometric elements. The corresponding time-domain DE is given as:

$$\frac{1}{c}\frac{\partial \Phi(\vec{r},t)}{\partial t} = \nabla\cdot D(\vec{r})\nabla\Phi(\vec{r},t) - \mu_a(\vec{r})\Phi(\vec{r},t) + S(\vec{r},t) \quad (2)$$

Where the temporal variation of photon fluence rate $\Phi(\vec{r},t)$ is modeled as the combination of photon diffusion, light absorption, and a source term. In a heterogeneous setting, the diffusion coefficient $D(\vec{r})$ is dependent on $\mu_a(\vec{r})$ and the reduced scattering coefficient $\mu_s'(\vec{r})$, given by $D(\vec{r}) = 1/[3(\mu_a(\vec{r}) + \mu_s'(\vec{r}))]$. The Robin-type boundary condition is typically assumed:

$$\Phi(\vec{r},t) + 2AD(\vec{r})\hat{n}\cdot\nabla\Phi(\vec{r},t) = 0 \quad (3)$$

where $A$ is the extrapolation length coefficient, determined by the refractive index mismatch at the boundary, and $\hat{n}$ is the outward-pointing normal.

The application of DE is subject to certain limitations. Specifically, it assumes that the $\mu_s \ll \mu_s'$ (Wang and Wu, 2007), and that the detectors are located at least one TMFP away from the light source (Jacques and Pogue, 2008). Since DE originates from the P$_1$ approximation to RTE, it assumes that the radiance is nearly isotropic, thereby losing accuracy in regions with strong directional photon transport, such as near sources, boundaries, or under oblique incidence.

In this study, we employ NIRFASTer (Dehghani, Dehghani et al., 2009) and Toast++ (Schweiger and Arridge, Schweiger and Arridge, 2014), both being widely used open-source FEM-based forward modeling toolkits in the DOI community. Their physical foundation is DE, typically solved under Robin boundary conditions. Both simulators feature a highly modular design and support parallel computing on the CPU through multithreading, providing efficient FEM computations. NIRFASTer integrates image processing (3DSlicer), mesh generation, and light propagation modelling in MATLAB, enabling comprehensive simulation and image reconstruction for various DOI modalities (Musgrove et al., 2007, Mellors and Dehghani, 2021), in particular CW/TD-DOT (Jiang et al., 2021, Althobaiti et al., 2017, Reisman et al., 2017). The latest version of NIRFASTer supports multithreaded acceleration with a GPU (Doulgerakis et al., 2017). Toast++, on the other hand, provides a rich set of low-level interfaces, making it well-suited for rapid model optimization and re-development in various research contexts (Ren et al., 2019). Compared to NIRFASTer, Toast++ offers greater flexibility in adjusting some basic settings of forward modelling, such as numerical schemes and boundary condition types.

Additionally, we assessed MMC (v2021.2) (Fang, 2010, Fang), a fast MC simulator designed to address previous limitations of MC simulations in complex geometries of the simulated object and low computational efficiency. MMC uses polyhedral meshes and Plücker coordinates to represent complex structures accurately, thereby improving geometric modeling precision. Furthermore, MMC supports CPU parallelization via OpenMP or SSE and GPU acceleration via

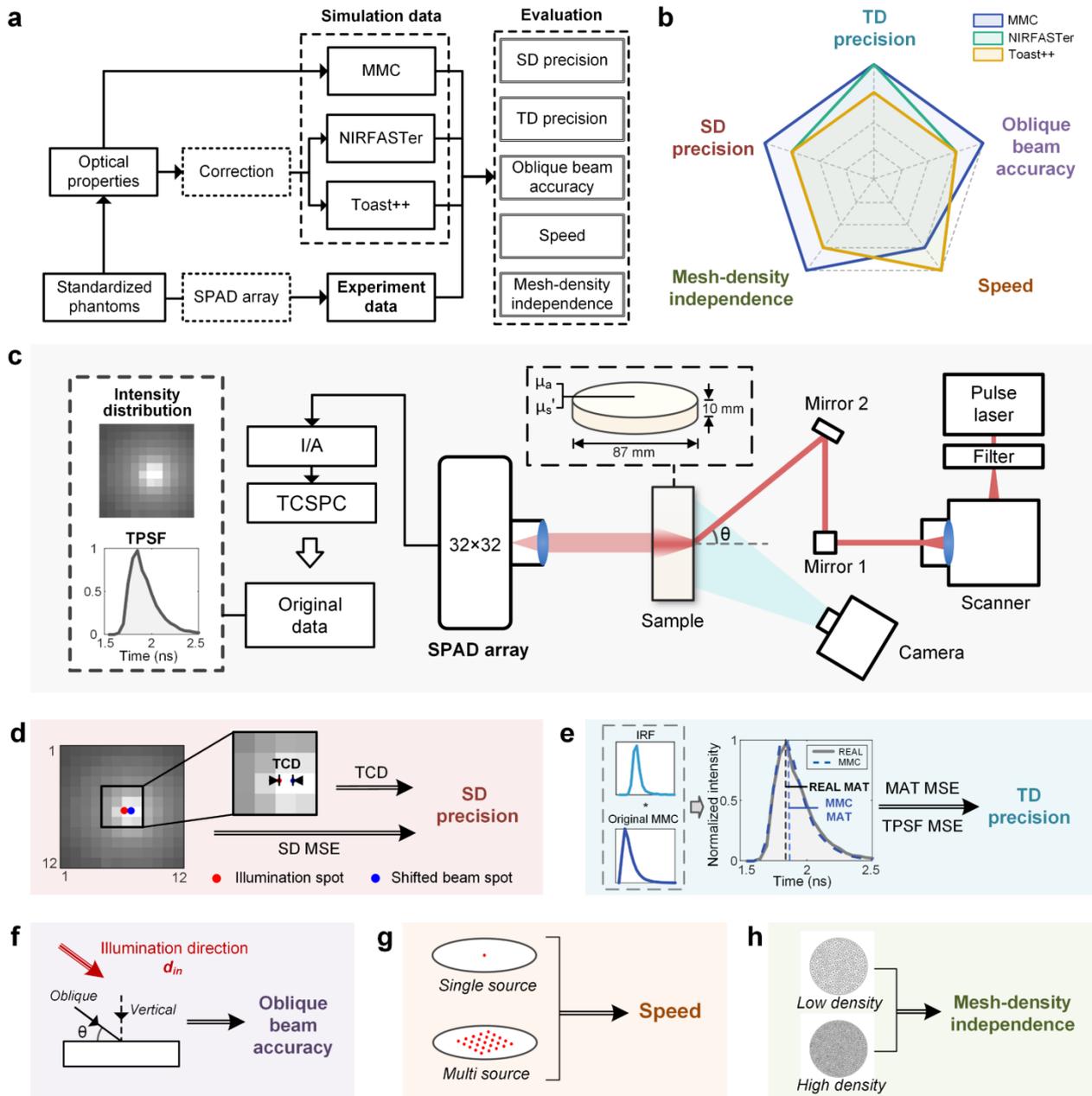

**Figure 1. The framework of the systematic evaluation of three light propagation simulators and the experimental setup.** (a) The validation workflow of three light simulators (MMC, NIRFASTer, and Toast++). Experimental measurements based on a TR SPAD array and standardized MEDPHOT phantoms were compared with simulation results on five different metrics. (b) The radar chart for comparing 5 metrics of three simulators, i.e., spatial-domain (SD) precision, time-domain (TD) precision, oblique beam accuracy, speed, and mesh-density independence. (c) The TR imaging system is equipped with a 32 × 32 SPAD array and a TCSPC module. (d-h) Detailed descriptions of each metric.

OpenCL, thereby achieving a balance between simulation accuracy and computational efficiency (Fang and Yan, 2019).

*2.2 Validation workflow and evaluation metrics*

The overall workflow of the evaluation is illustrated in figure 1a, where TR experimental data were directly compared with simulation data resulting from three simulators (MMC, NIRFASTer, and Toast++). NIRFASTer and Toast++ were executed on the CPU with parallelization, while MMC employed GPU acceleration through OpenCL. The experimental data were acquired from a TD DOI system based on a SPAD array applied to standard homogeneous phantoms following a MEDPHOT protocol (Pifferi et al., 2005). Prior to simulation, the optical properties of the phantoms ($\mu_a$ and $\mu_s'$) were assigned based on the given values of each standard phantom in real experiments. For both FEM-based simulators, the virtual source correction technique was used to compensate the oblique light illumination. In contrast, MMC adopts direct input of the actual incident angle without correction. Once both the experimental and simulated datasets were obtained, the performance of these simulators was systematically evaluated across five metrics, namely SD precision, TD precision, oblique beam accuracy, speed, and mesh-density independence. The overall evaluation results were summarized as a single radar chart, visualizing the advantages and disadvantages of each metric (figure 1b). The detailed explanation of each metric is illustrated in figure 1d-h.

1) SD precision: The accuracy of light distribution in SD is relevant to energy deposition in the detector plane and provides the primary information for many DOI reconstruction tasks. The overall metric of SD precision consists of two components: transmission center displacement (TCD) and SD mean squared error (MSE). The former is defined as the Euclidean distance between the statistically estimated center of the transmitted photon distribution $(x_t, y_t)$ and the illumination source location $(x_0, y_0)$, given by:

$$TCD = \sqrt{(x_t - x_0)^2 + (y_t - y_0)^2} \quad (4)$$

Here $(x_t, y_t)$ is estimated by solving the following optimization problem:

$$\min_{A,C,x_t,y_t,\sigma_x,\sigma_y} \sum_{h=1}^{H}\sum_{w=1}^{W}\left[A \cdot exp(\delta) + C - \sum_{t=1}^{T} I(x_h, y_w, t)\right]^2 \quad (5)$$

Here, $\delta = -\left(\frac{(x_t-x_h)^2}{2\sigma_x^2} + \frac{(y_t-y_w)^2}{2\sigma_y^2}\right)$. $(x_h, y_w)$ represents the real-world coordinate at the image pixel $(h, w)$. $(x_t, y_t)$ can be obtained by fitting a 2D Gaussian model with least-square measures, and $I(x_h, y_w, t)$ is the measured intensity at time index t at $(x_h, y_w)$. H and W denote the maximum height and width of the pixels in the detected region. T represents the total number of time steps. $\sigma_x^2$ and $\sigma_y^2$ represent the variances of the Gaussian profile along the $x$ and $y$ directions, respectively. $A$ is the amplitude of the Gaussian distribution, and $C$ is an offset constant accounting for the baseline intensity.

SD MSE is defined as the normalized MSE between the two-dimensional integrated photon intensity over time in simulation and measurement:

$$MSE_{SD} = \frac{1}{HW}\sum_{h=1}^{H}\sum_{w=1}^{W}\left(I_{SD}^{simu}(h,w) - I_{SD}^{real}(h,w)\right)^2 \quad (6)$$

where $I_{SD}^{simu}(h,w)$ and $I_{SD}^{real}(h,w)$ denote the normalized integrated photon intensity images of simulations and real experiments (Schweiger and Arridge, 1999), respectively, given by:

$$I_{SD}(h,w) = \frac{\sum_{t=1}^{T} I(h,w,t)}{max \sum_{t=1}^{T} I(h,w,t)} \quad (7)$$

In the end, the overall score of SD precision is determined by the mean value of TCD and $MSE_{SD}$ after normalization by their maximum values:

$$P_{SD} = \frac{1}{2}\left(T\tilde{C}D + M\tilde{S}E_{SD}\right) \quad (8)$$

2) TD precision: The temporal information represented by the ToF histogram contains rich information for disentangling the mixed scattering and absorbed photons in depth (Farina et al., 2017). The TD precision is evaluated by two components: the MSE of mean arrival time (MAT) and the MSE of temporal point spread function (TPSF), which are denoted as $MSE_{MAT}$ and $MSE_{TPSF}$ respectively. The former component $MSE_{MAT}$ is measured by the normalized MSE of photon arrival times across all pixels, given by:

$$MSE_{MAT} = \frac{1}{HW}\sum_{h=1}^{H}\sum_{w=1}^{W}\left(I_{MAT}^{simu}(h,w) - I_{MAT}^{real}(h,w)\right)^2 \quad (9)$$

where $I_{MAT}(h,w)$ is the first-order temporal moment, which is frequently used in TD-DOT (Schweiger and Arridge, 1999), given by:

$$I_{MAT}(h,w) = \frac{\sum_{k=1}^{T} t_k \cdot I(h,w,t_k)}{\sum_{k=1}^{T} I(h,w,t_k)} \quad (10)$$

where $t_k$ represents the k-th discrete time.

The other term $MSE_{TPSF}$ is the normalized MSE between the TPSFs at all pixels of simulation and measurement, retaining the complete original temporal information, which is potentially important for DOI reconstruction using full ToF histogram information (Gao et al., 2008). $MSE_{TPSF}$ is calculated by:

$$MSE_{TPSF} = \frac{1}{HWT}\sum_{h=1}^{H}\sum_{w=1}^{W}\sum_{t=1}^{T}\left(I^{simu}(h,w,t) - I^{real}(h,w,t)\right)^2 \quad (11)$$

Similar to SD precision, the overall score of TD precision $P_{TD}$ is determined by the mean value of normalized $MSE_{MAT}$ and $MSE_{TPSF}$:

$$P_{TD} = \frac{1}{2}\left(M\tilde{S}E_{MAT} + M\tilde{S}E_{TPSF}\right) \quad (12)$$

3) Oblique beam accuracy: the non-contact DOI systems may encounter free-space illumination with an oblique incidence beam, in case the object surface is uneven. The accuracy in simulating the oblique incidence beam is evaluated mainly for FEM simulations, where the incident angle cannot be directly simulated due to the intrinsic limitations of DE. The oblique beam accuracy (OBE) is determined by measuring the overall reduction in $MSE_{SD}$ and $MSE_{TPSF}$ achieved through incorporating the actual oblique incident angle into the simulation.

$$OBE = \frac{1}{2}\left(\frac{MSE_{SD}^{dir} - MSE_{SD}^{cor}}{MSE_{SD}^{dir}} + \frac{MSE_{TPSF}^{dir} - MSE_{TPSF}^{cor}}{MSE_{TPSF}^{dir}}\right) \quad (13)$$

where the superscript of *dir* refers to the direct simulation assuming the perpendicular incidence to the surface, while *cor*

denotes the angular corrected simulation, reflecting the improvement achieved by accounting for the incident angle.

4) Speed: Simulation speed reflects the computational efficiency of the adopted method, which is highly important for TD DOI systems containing rich temporal information. In many cases, numerous illumination points can be applied (Gao et al., 2022, Zhao et al., 2021). Therefore, the average time cost for both single and multiple point sources (source number = 1, 25, 100) is calculated:

$$Speed = \frac{3}{\tilde{T}_1 + \tilde{T}_{25} + \tilde{T}_{100}} \quad (14)$$

where $\tilde{T}_n$ is the normalized time cost of the simulation using n sources for each simulator.

5) Mesh-density independence: A higher mesh density can enhance simulation accuracy, but in turn requires increased computational time for both mesh generation and the subsequent simulation. Mesh-density independence (MDI) reflects the change of the $MSE_{TPSF}$ values and time cost under different levels of mesh density, given by:

$$MDI_{MSE} = \frac{MSE_{TPSF}(coarse\ mesh) - MSE_{TPSF}(dense\ mesh)}{MSE_{TPSF}(coarse\ mesh)} \quad (15)$$

and

$$MDI_{time} = \frac{T(dense\ mesh) - T(coarse\ mesh)}{T(dense\ mesh)} \quad (16)$$

Here, the level of mesh density is controlled by "maxvol" denoting the maximum allowable volume (mm$^3$) of a tetrahedral element, which is used in a mesh generator iso2mesh (Fang). In our comparison, we chose a coarse mesh and a dense mesh with maxvol = 5 and 0.2 mm$^3$, respectively. $T$ denotes the time cost of the specific maxvol value. Lastly, the overall score of mesh-density independence is determined by the average of $MDI_{MSE}$ and $MDI_{time}$.

*2.3 Experimental Setup*

The TR DOI system is configured in a transmission geometry (figure 1c). The illumination module comprises a wavelength-tunable picosecond pulsed laser with a repetition rate of 78 MHz (SuperK EXTREME, NKT, Denmark), an acousto-optic tunable filter (VARIA, NKT, Denmark), a galvanometric scanner for controlling the beam position, and two mirrors for controlling the incident angle, which is employed under large-angle oblique incidence conditions. The detection module mainly consists of a customized 32 × 32 SPAD array (Zhang et al., 2018) equipped with a 15 mm focus lens (ML183M, Theia Technologies, USA). Integrated with Time-Correlated Single Photon Counting (TCSPC) circuitry, the SPAD array records the photon counts over multiple excitation cycles to generate a ToF histogram for each pixel with a minimum time bin size of 50 ps (Zhang et al., 2018). The excitation wavelength and intensity were precisely configured, and custom acquisition software was developed to enable the SPAD array to perform repeated signal acquisitions, allowing accurate fitting of TPSF. The laser and SPAD array were synchronized using a timing controller, with the laser providing the trigger signal for TCSPC. A mechanical holder was used to stabilize the phantom and minimize ambient light interference, ensuring experimental consistency. The field-of-view (FOV) of the system is set to approximately 60 mm × 60 mm, finetuned by using a motorized lift to align the phantom surface with the detection window. For all phantom experiments, the wavelength of the laser was tuned to 780 nm with a spectral bandwidth of 10 nm. A galvanometric scanner and mirrors were used to adjust the position and direction of the incident beam, ensuring that it entered the center of the FOV.

*2.4 Phantom experiments*

All experiments were performed on 16 standard homogeneous phantoms following a MEDPHOT protocol (BioPixS0087, BioPIXS, Ireland), all with the same cylindrical geometry (radius = 43.5 mm, height = 10 mm, figure 2a). The set of phantoms covers 4 levels of absorption (labelled as 1-4) and 4 levels of scattering (labelled as A-D), both ranging from low to high and measured at 780 nm (Inc.), with specific values provided in Table 1 and plotted in figure 2b.

During each measurement, the pulsed laser was directed onto the phantom surface at two different incident angles (θ = 2° and 58°). The TR signals were recorded by the SPAD array, resulting in 256 temporal bins in the raw data. The impulse response function (IRF) was measured under identical conditions by scanning a dense grid of illumination points across the detection area. For each point, the SPAD array recorded the photon arrival histogram, and the TPSF was extracted from the pixel with the strongest signal. The IRF was constructed as the collection of TPSFs obtained from all scanned positions and used for subsequent data analysis. Prior to comparison with simulation data, the measured raw data underwent systematic preprocessing, including defective pixel correction, spatial and temporal windowing, and normalization. To address possible malfunctioning or abnormal responses in individual SPAD pixels, the value of each defective pixel was linearly interpolated using eight neighbouring pixels, enhancing spatial data completeness and continuity. To enhance the effective signal and suppress background noise, a central 12 × 12 pixel region with sufficiently high photon counts was extracted for SD analysis, where each pixel corresponds to an area of approximately 2 mm × 2 mm. For TD analysis, a window of 20 time bins centered around the signal peak was selected in order to achieve a part exceeding 1% of the peak intensity which will be analyzed later, effectively eliminating low-intensity noise and improving the signal-to-noise ratio (SNR). All TPSFs were convolved with the IRF over 256 time bins for each pixel to account for system response artifacts.

**Table 1.** Optical properties of standard phantoms.

| Phantom | μ$_a$ (cm$^{-1}$) | μ$_s$' (cm$^{-1}$) | Phantom | μ$_a$ (cm$^{-1}$) | μ$_s$' (cm$^{-1}$) |
|---|---|---|---|---|---|
| A1 | 0.05 | 4.9 | C1 | 0.05 | 14.0 |
| A2 | 0.10 | 5.2 | C2 | 0.10 | 13.5 |
| A3 | 0.22 | 5.4 | C3 | 0.20 | 14.0 |
| A4 | 0.33 | 5.5 | C4 | 0.32 | 14.3 |
| B1 | 0.05 | 9.2 | D1 | 0.05 | 18.2 |
| B2 | 0.10 | 9.2 | D2 | 0.10 | 18.2 |
| B3 | 0.21 | 9.6 | D3 | 0.21 | 18.8 |
| B4 | 0.32 | 10.1 | D4 | 0.33 | 19.6 |

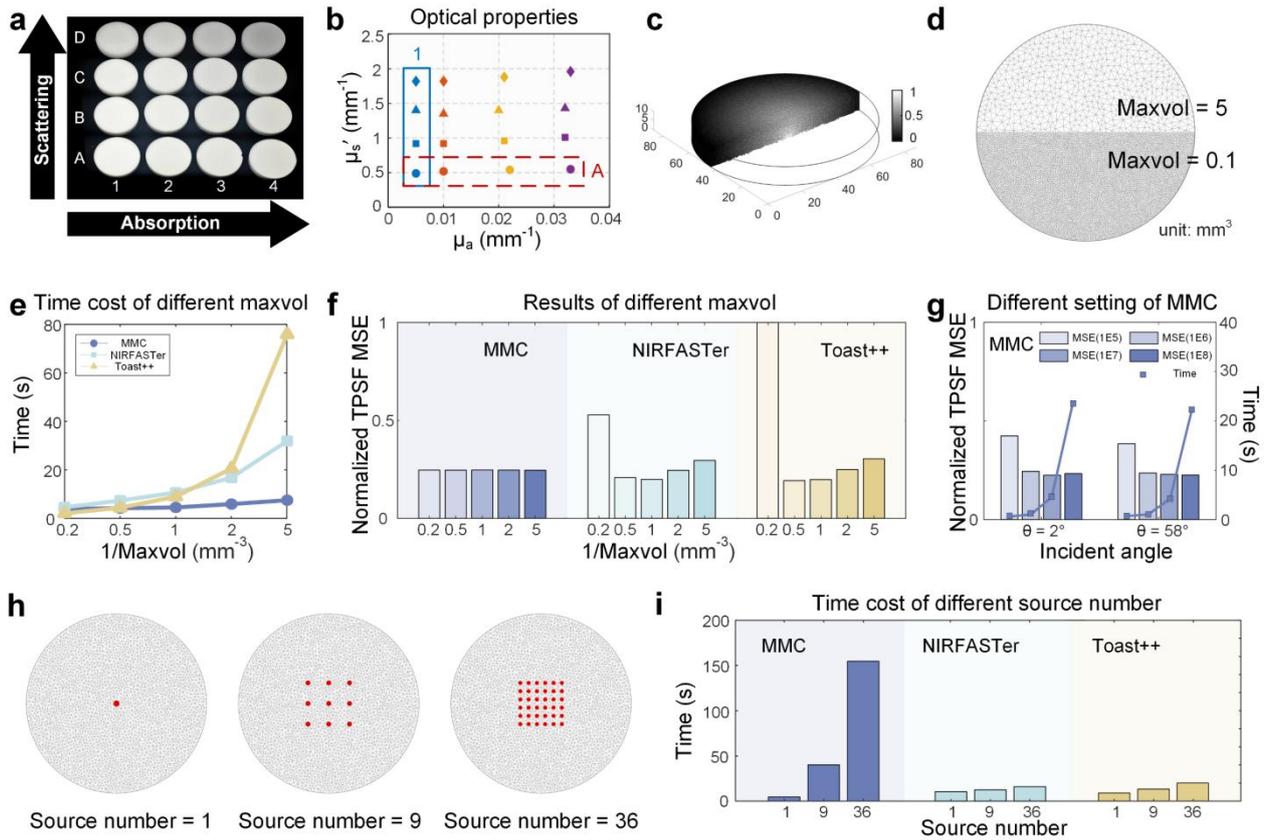

**Figure 2. Evaluation of simulation accuracy, runtime efficiency, and configuration effects in homogeneous phantom experiments.** (a) The photograph of 16 homogeneous phantoms featuring different optical properties. (b) The ground truth (GT) values of $\mu_a$ and $\mu_s'$ of these phantoms. (c) An example of light propagation modeling with MMC based on a 3D mesh. (d) Top view of mesh examples with different levels of density controlled maximum allowable volume of a tetrahedral element (maxvol = 5 and 0.1 mm$^3$). (e) Simulation runtimes for three simulators under different mesh density levels. (f) Simulation accuracy measured by TPSF MSE with varying mesh density levels for three simulators under a 50 ps time step. (g) Simulation accuracy measured by TPSF MSE and runtimes under varying launched photon counts in the MMC simulator. (h) Simulation cases using different numbers of light sources with single dot, 3 × 3 grid, and 6 × 6 grid. (i) Runtime comparison across simulators using different numbers of light sources.

## 2.5 Numerical simulations and optimization of simulation configurations

All simulators in this study are mesh-based, which requires a uniform mesh to ensure consistent comparison across different simulators. A 3D cylindrical mesh consistent with the physical dimensions of the standard phantom was generated using the iso2mesh toolbox (Fang, Tran et al., 2020) (figure 2c). The mesh consists of 21245 nodes and 113382 tetrahedral elements, each assigned with certain optical properties ($\mu_a$ and $\mu_s'$) and refractive index (n, default value of 1.4). The light source was modeled as a point source. To ensure simulation efficiency, a duration of 1.6 ns was simulated, as the effective measured data is primarily concentrated within this time window. All simulations were performed on a workstation equipped with an Intel i7-14700 CPU (20 cores (28 threads), base frequency = ~2.1 GHz, RAM = 32GB) and an NVIDIA GeForce RTX 3060 GPU.

For ensuring a fair and efficient comparison with experimental results, we optimized the simulation configurations on the aspects of mesh density, time step (both for all simulators), and launched photon number (for MMC only). Firstly, during the meshing stage, the mesh density is controlled by the maxvol, i.e., a small value of maxvol yields a high-density mesh and vice versa (figure 2d). We evaluated the modelling accuracy and computational cost for 5 levels of mesh density (maxvol = 0.2, 0.5, 1, 2, and 5 mm$^3$). Next, we adapted the value of the time step by measuring the stability of the TPSF profile of FEM simulators. Hence, 3 values of time step (50 ps, 25 ps, and 10 ps) were tested to seek the optimal value. For TD FEM simulations, the Crank-Nicholson scheme is applied, isotropic sources were employed to match the implemented source correction in this study, and the detector readings were obtained by applying Gaussian weighting within a predefined radius around each detection point. Finally, different numbers of simulated photons were launched in MMC, ranging from $10^5$ to $10^8$, considering that the photon number directly impacts the accuracy of MC simulations. During the optimization procedure, all simulations were performed on phantom C4, a representative phantom ($\mu_a$ = 0.032 mm$^{-1}$, $\mu_s'$ = 1.43 mm$^{-1}$ ) mimicking the optical characteristics of human skin (Bashkatov et al., 2005, Kono and Yamada, 2019).

It is worth noting that our study primarily focuses on the single-point source illumination case for all five metrics except the simulation speed. For complex imaging applications, such as any form of optical tomography or SFDI, point-by-point raster scanning or structured-pattern illumination are often used. Therefore, we also incorporated a comparison of simulation speed for single versus multiple light sources under various methods (Equation (14), figure 2h).

### 2.6 Virtual source correction for FEM simulators

Unlike MC methods, the FEM solution based on the DE neglects the anisotropy of light propagation. For FEM simulators, an incidence beam perpendicular to the object surface is typically assumed. Such a setting holds true for fiber-based contact systems such as fNIRS (Ferrari and Quaresima, 2012). Nevertheless, this may bring significant errors for oblique light illumination in a free-space imaging system such as that in small animal imaging and fluorescence-guided surgery (Daly et al., 2019). To address this, we incorporate virtual source correction to account for oblique light illumination (Wang and Jacques, 1995). More specifically, the real light source position $(x, y, z)$ is relocated inward to a new virtual position $(x + \Delta x, y + \Delta y, z + \Delta z)$ within the medium, effectively accounting for the directionality of the incident light with a direction vector $d_{in} = (a, b, c)$ (figure 3a). The virtual source is displaced with $\Delta x$, $\Delta y$, and $\Delta z$, given by:

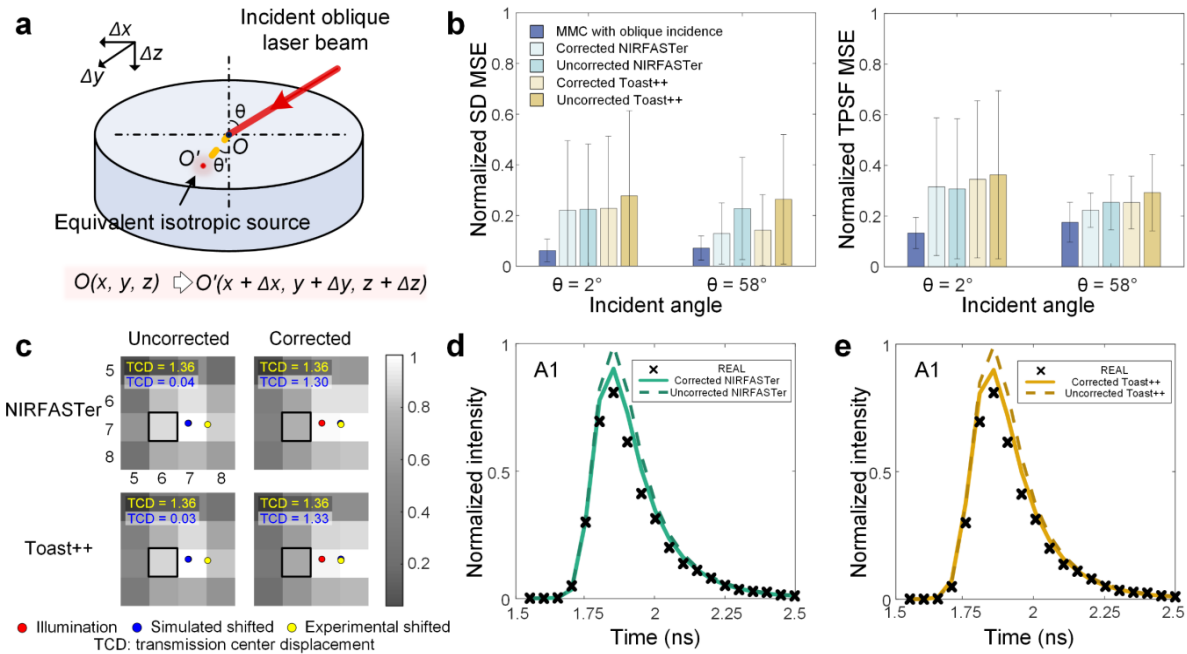

**Figure 3. Evaluation of source correction for oblique incidence in FEM simulations. (a) Source correction for the oblique incident beam.** (The illumination position on the surface, denoted as *O*, is located at (*x, y, z*). In FEM simulation, the corrected virtual source is relocated at *O'* at a new position (*x + Δx, y +Δy, z + Δz*). (b) $MSE_{SD}$ and $MSE_{TPSF}$ of simulators with and without source correction for oblique incident beam with θ = 2° and θ = 58°, (c) Intensity distribution of FEM simulations with and without source correction. (d) Comparison of off-center TPSF profiles in experimental measurements and NIRFASTer simulation with and without source correction at the selected pixel (7, 6) for θ = 58°, (e) Comparison of off-center TPSF profiles in experimental measurements and Toast++ simulation with and without source correction at the selected pixel (7, 6) for θ = 58°.

**Table 2.** Comparison of the computation time (unit: s) under varying simulation configurations.

| Maxvol | Time step | | | | | | Photon number | | | |
|---|---|---|---|---|---|---|---|---|---|---|
| | 50 ps | | 25 ps | | 10 ps | | $10^5$ | $10^6$ | $10^7$ | $10^8$ |
| | NIR[a] | TOA[b] | NIR | TOA | NIR | TOA | MMC | MMC | MMC | MMC |
| 5 | 3.02 | 1.51 | 4.57 | 2.08 | 9.28 | 3.76 | 0.50 | 0.90 | 3.91 | 22.00 |
| 2 | 4.49 | 3.10 | 7.36 | 4.38 | 15.64 | 8.26 | 0.70 | 0.94 | 4.18 | 22.08 |
| 1 | 6.27 | 6.30 | 10.62 | 8.93 | 23.40 | 18.19 | 0.95 | 1.13 | 4.51 | 23.40 |
| 0.5 | 9.33 | 15.52 | 16.80 | 20.58 | 36.88 | 42.50 | 1.38 | 1.57 | 5.89 | 26.05 |
| 0.2 | 17.67 | 44.27 | 31.91 | 75.84 | 72.16 | 167.07 | 2.89 | 3.15 | 7.48 | 31.28 |

[a]NIR, NIRFASTer; [b]TOA, Toast++.

$$(\Delta x, \Delta y, \Delta z) = \left( \frac{x}{\|d_{in}\| \cdot n \cdot \mu_t'}, \frac{y}{\|d_{in}\| \cdot n \cdot \mu_t'}, \frac{z}{\|d_{in}\| \cdot n \cdot \mu_t'} \right) \quad (16)$$

Where $\|d_{in}\| = \sqrt{a^2 + b^2 + c^2}$ and the total attenuation coefficient $\mu_t' = 0.35\mu_a + \mu_s'$.

## 3. Results

### 3.1 Optimization of simulation configuration

To pursue efficient and comparable validation in an identical condition for different simulators, we first optimized several important configuration parameters, which mainly contain mesh density, time step, and the launched photon number (only for MMC). The time cost is recorded in Table 2. Regarding meshing, significant differences in both computational time and accuracy were observed among the three simulators under varying levels of mesh density, averaged over the results from both incident angles under 25 ps. Figure 2e shows that the time cost for Toast++ and NIRFASTer increases substantially with mesh density, being about 35 and 6 times higher, respectively, when using the densest mesh compared to the coarsest one. In contrast, MMC requires less time overall, averaging 3.6 s with only ~1.6 s variation across different mesh densities. This preliminary result shows that the time cost of Toast++ is more strongly affected by mesh complexity than NIRFASTer and MMC. In terms of accuracy, FEM-based simulators exhibit a significant reduction in error as mesh density increases from maxvol = 5 mm³ to maxvol = 1 mm³, whereas MMC shows minimal change in error across different mesh densities, indicating weaker dependence on mesh granularity (figure 2f). Besides, the set of 25 ps or less time steps could achieve a stable and smooth original TPSF profile, which is needed to avoid the error caused by poor quality simulation (figure S1). To balance both accuracy and computational efficiency, maxvol = 1 mm³ (corresponding to a mesh with 21245 nodes) and time step = 25 ps were used as the optimal setting for all subsequent simulations.

In parallel, the impact of photon counting number on the MMC results was also evaluated. Figure 2g demonstrates that the simulation accuracy measured by MSE plateaus when the number of launched photons reaches $10^7$ or higher, while further increasing the photon number significantly prolongs the simulation time with additional computational load. When the photon count is $10^7$, the value of the normalized $MSE_{TPSF}$ is 0.5168, and the computational time is 4.51 s. However, for the photon count of $10^8$, the normalized $MSE_{TPSF}$ increases slightly to 0.5222, and the computational time is 23.4 s with a significant rise of 419%. Therefore, photon count = $10^7$ was adopted as a trade-off choice between accuracy and efficiency.

Beyond the meshing condition and simulated photon number, the influence of light source quantity on the runtime was also assessed (figure 2h). Figure 2i shows that the simulation time of MMC increases remarkably with the number of point sources, significantly exceeding that of the other two FEM simulators. More detailed comparison of the computational time under different simulation configurations is given in Table. 2.

### 3.2 Source correction for oblique incident light

We evaluated the effect of incorporating incident angle information in forward modelling. Figure 3b and 3c show that MMC achieves the lowest $MSE_{SD}$ and $MSE_{TPSF}$ in all cases of oblique incidence illumination. For a large-angle incidence illumination (θ = 58°), the mean values of $MSE_{SD}/MSE_{TPSF}$ for NIRFASTer and Toast++ simulation after correction are 0.156/0.031 (average drop: 56.3%/12.3%) and 0.180/0.039 (average drop: 57.0%/13.4%), respectively. For small-angle incidence (θ = 2°), the impact of source correction is less. In this case, there is a relatively average small difference below 2%/11.4% for NIRFASTer/Toast++. This result suggests that angular corrections are less influential under small-angle incidence conditions. Excluding the small angular conditions, the SD and TD results show that virtual source correction reduces modelling error by at least 34% on average.

Besides statistical analysis, we also showcase the individual results from an exemplar phantom (A1, $\mu_a$ = 0.005 mm$^{-1}$, $\mu_s'$ = 0.49 mm$^{-1}$) illuminated by oblique incident light with an angle θ = 58° (figure 3c-e). We compare the values of TCD (Equation (4)) and the TPSF of a selected pixel before and after correction. Figure 3c shows that the corrected model generates a more precise centroid shift of the simulated photon intensity distribution (blue dot), which is closer to that of the measurement data (yellow dot). The position of the illumination beam is noted as the red dot. Comparing the TCD values estimated from simulation and experimental data, the difference (denoted as ΔTCD) was 1.30 mm when NIRFAST was used for simulation and 1.33 mm when Toast++ was used. After virtual source correction, the TCD values reduce to 0.04 for NIRFASTer and 0.03 for Toast++, both showing better agreement with the experimental measurements, with a significant decrease of 97% and 98%. At the pixel location (7, 6), the TPSFs from the corrected model more closely match the experimental data for both simulators (figure 3d and 3e). Source correction also improves the accuracy of the peak amplitude and full width at half maximum (FWHM) of the individual simulated TPSF curves. Up to now, we have demonstrated that virtual source correction in FEM can effectively improve the accuracy in both spatial and time domains, thus enhancing overall simulation fidelity. In the following analysis, we adopted the virtual source correction in all FEM simulation studies.

### 3.3 Spatial-domain precision analysis

To evaluate the SD precision of different simulators, metrics of TCD and $MSE_{SD}$ were calculated for varying optical properties by grouping the phantoms according to the same $\mu_a$ or $\mu_s'$. As discussed in section 3.2, for small-angle incidence with e.g., θ ≤ 10°, TCD is relatively small and negligible (Kienle et al., 1996). Hence, the analysis of TCD only focuses on the large-angle incidence with θ = 58°, where the illumination spot shift becomes predominant. All TCD differences between simulations and measurements (ΔTCD) are less than 0.8 mm, indicating a high level of agreement and confirming the spatial accuracy of the simulated photon distribution (figure 4a). Among the three simulators, MMC achieves the lowest ΔTCD, outperforming two FEM simulators, as observed by the data grouped either by $\mu_a$ or $\mu_s'$. Using $MSE_{SD}$ as a quantitative metric, the three simulators exhibit distinct patterns in applicability and stability according to the optical property-based grouping (figure 4b). Both FEM simulators show slightly increased $MSE_{SD}$ when the optical absorption increases, whereas the trend is opposite, showing significant decrease of $MSE_{SD}$ when the scattering increases. More specifically, Toast++ generates the largest $MSE_{SD}$ for a low scattering case (Group A), but this error drops significantly at a higher scattering level (Group D). A similar trend is observed for NIRFASTer, although the overall errors are comparatively lower. Compared to FEM simulators, MMC exhibited superior stability and accuracy with consistently low $MSE_{SD}$ values, which is expected because it does not rely on the diffusion approximation. This difference is especially relevant in the low-scattering regime where the diffusion approximation is less valid. A statistical analysis of all results,

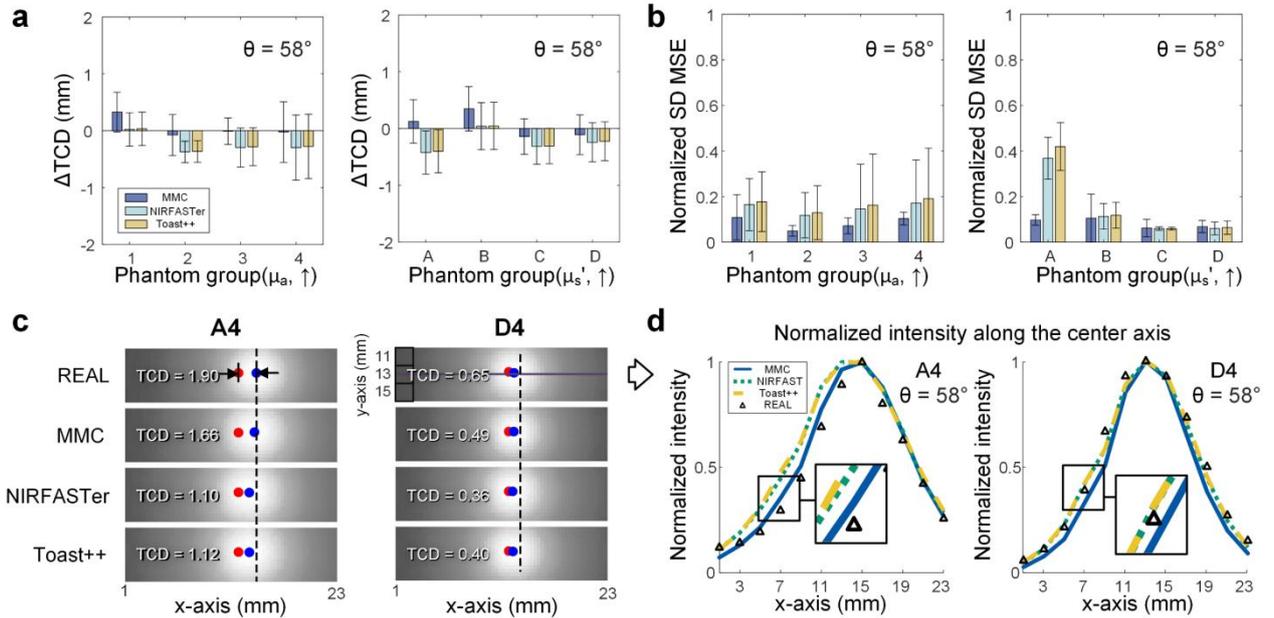

**Figure 4. Quantitative analysis of SD precision using TCD and $MSE_{SD}$ for varying optical parameters.** (a) The difference of TCD between measurements and simulation results (ΔTCD) achieved from different simulators for grouped phantoms with the same $\mu_a$ or $\mu_s'$ under large-angle oblique incidence illumination. (b) Comparison of $MSE_{SD}$ for phantoms with the same $\mu_a$ or $\mu_s'$ for different incidence angles. (c) A cropped detection window (2 × 12 pixels) labeled with the illumination position (blue dot), the estimated center of the transmitted photons (red dot), and TCD value. (d) Normalized intensity profiles between simulation and measurement for phantoms A4 and D4 with θ = 58°. The profiles are drawn along the dashed lines in (c).

including mean MSE values and the standard deviations, is summarized in Table 3.

Besides the statistical analysis, we take a closer look at illustrative data of phantom A4 ($\mu_a = 0.033$ mm$^{-1}$, $\mu_s' = 0.55$ mm$^{-1}$) and phantom D4 ($\mu_a = 0.033$ mm$^{-1}$, $\mu_s' = 1.96$ mm$^{-1}$) featuring different levels of light scattering and absorption. A cropped detection window sizing 3 × 12 pixels is visualized in figure 4c. The illumination position (blue dot), the estimated center of the transmitted photons (red dot), and the TCD value are shown for real measurements and simulators. For low scattering objects like A4 with large-angle oblique illumination, the TCD value is typically large, reflected in both real measurement and simulation (TCD = 1.90, 1.66, 1.10, 1.12 mm for measurement, MMC, NIRFASTer, and Toast++, respectively) because of the larger scattering length. In contrast, for heavily scattering objects like D4, the TCD value is much smaller (TCD = 0.65, 0.49, 0.36, 0.40 mm for measurement, MMC, NIRFASTer, and Toast++). An intensity profile along the center axis further reveals these distinctions (figure 4d). This indicates that photon propagation becomes more diffuse and less directional in highly diffuse media, in which $\mu_s'$ appear large.

**TABLE 3.** Accuracy evaluation of different simulators via MSE analysis (mean ± std)

|         | MMC               | NIRFASTer       | Toast++         |
|---------|-------------------|-----------------|-----------------|
| SD MSE  | **0.072 ± 0.053** | 0.257 ± 0.321   | 0.267 ± 0.332   |
| MAT MSE | **0.226 ± 0.133** | 0.333 ± 0.169   | 0.354 ± 0.175   |
| TPSF MSE| **0.179 ± 0.080** | 0.227 ± 0.069   | 0.258 ± 0.106   |

The overall score of SD precision can be measured by the proposed index $P_{SD}$ according to Eq. 7, yielding $P_{SD}$ = 0.590, 0.981, 0.989 for MMC, NIRFASTer, and Toast++, respectively. Both statistical analysis and illustrational phantom data suggest that MMC is more advantageous in both accuracy and stability than NIRFASTer and Toast++ for various combinations of $\mu_a$ and $\mu_s'$. The SD accuracy of NIRASTer and Toast++ is nearly equivalent.

### 3.4 Time-domain precision analysis

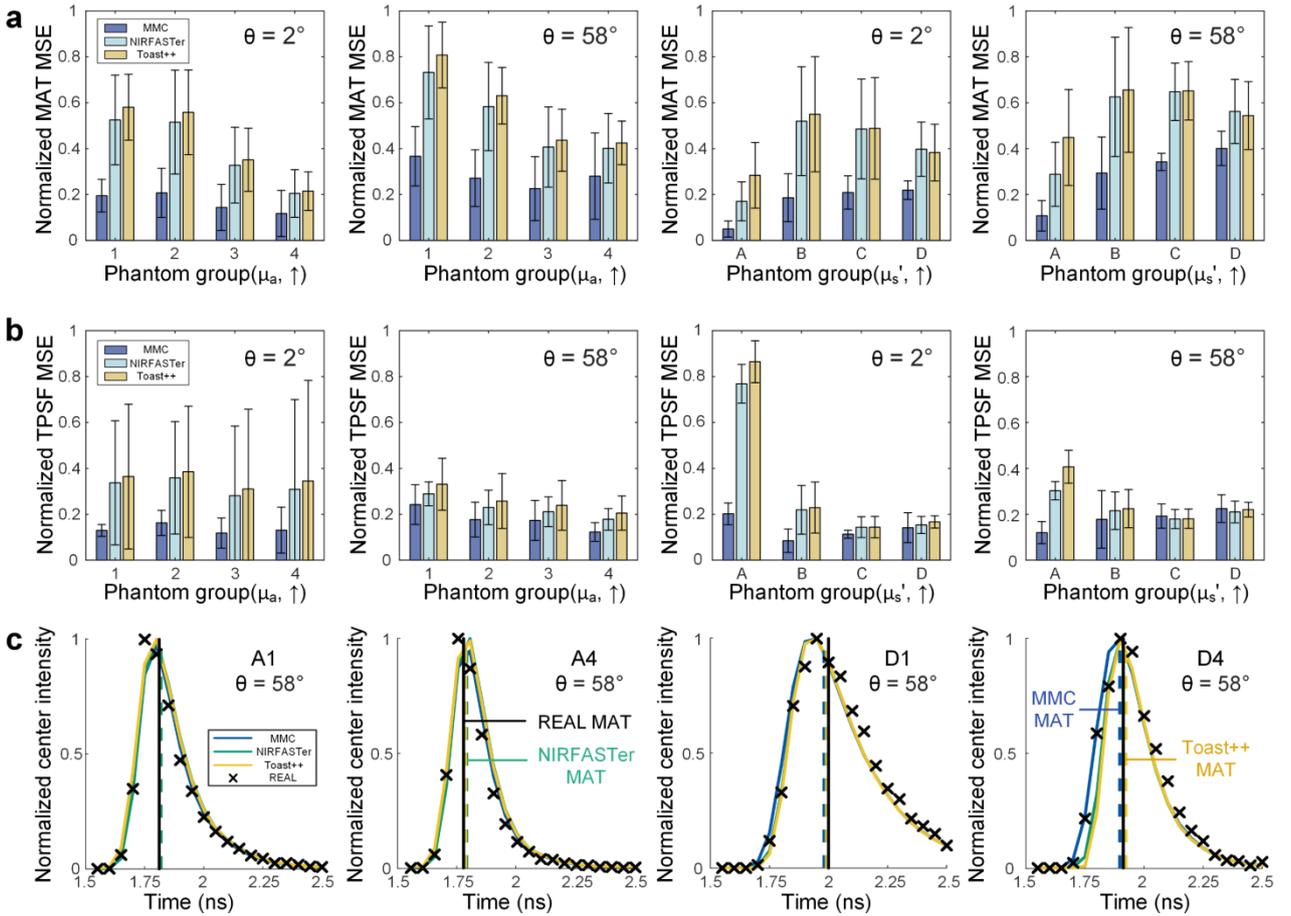

**Figure 5. Evaluation of TD precision based on $MSE_{MAT}$ and $MSE_{TPSF}$ for varying optical properties.** Comparison $MSE_{MAT}$ (a) and $MSE_{TPSF}$ (b) of different simulators for the grouped phantoms with the same $\mu_a$ or $\mu_s'$ with different incidence angles (θ = 2° and θ = 58°). (c) TPSF curves and MAT at the central pixel of the simulation results from phantom A1, A4, D1, and D4 with θ = 58°.

We further conducted comparison of the three simulators in TD by calculating $MSE_{MAT}$ and $MSE_{TPSF}$. Figure 5a shows the error between simulated and measured MAT. The $MSE_{MAT}$ values of all simulators decrease as $\mu_a$ increase. The correlation between $\mu_a$ and $MSE_{MAT}$ was calculated and averaged on both small and large incidence angles, which yields the following results: Pearson correlation coefficient = -0.307, -0.636, -0.769 for MMC, NIRFASTer and Toast++, respectively. It also suggests that the accuracy of photon arrival time calculated from the NIRFASTer and Toast++ is more sensitive to the change of light absorption. In contrast, $MSE_{MAT}$ values for all simulators are less sensitive to the change of light scattering (figure 5a). Among three simulators, MMC consistently exhibits the highest accuracy and stability for most conditions. Detailed $MSE_{MAT}$ values can be found in Table 3. We also analyzed the TPSFs across all pixels. The $MSE_{TPSF}$ exhibits a similar trend as the $MSE_{SD}$ in figure 4b, suggesting a high degree of correlation between spatial and temporal accuracy, which is important for reconstruction tasks involving ToF. Notably, both FEM based simulators showed a negative correlation with $\mu_s'$. However, slight differences were observed as the $MSE_{TPSF}$ generally exhibited a negative correlation with $\mu_a$ across the three simulators. $MSE_{TPSF}$-$\mu\_a$ Pearson correlation = -0.298, -0.299, -0.209, and $MSE_{TPSF}$-$\mu_s'$ Pearson correlation = -0.402, -0.672, -0.589 for MMC, NIRFASTer and Toast++, respectively. This indicates that higher $\mu_a$ and $\mu_s'$ generally lead to improved modeling accuracy within the tested range.

In addition to the statistical analysis, we showcase the TPSF profiles and the corresponding MAT (marked with vertical lines) for several representative phantoms (A1, A4, D1, and D4) in figure 5c, featuring the smallest or largest $\mu_a$ and $\mu_s'$ (see detailed values in Table 1). The FWHM of TPSF shows an evident dependence on simulation parameters, which decreases with increasing $\mu_a$ and increases with increasing $\mu_s'$. This indicates that under high-scattering conditions, photons undergo more scattering events, broadening the TPSF, whereas high absorption reduces the number of late-arriving photons, narrowing the TPSF. Notably, when FWHM is large, simulated photon arrival counts in the later time bins may underestimate the measured values, while they may be overestimated if FWHM is narrow, leading to increased errors, especially in the tail region of the TPSF.

The overall score of TD precision can be measured by the proposed index $P_{TD}$ according to Equation (12), which results in $P_{TD}$ = 0.667, 0.910 and 1 for MMC, NIRFASTer, and Toast++, respectively. Again, MMC achieves the highest accuracy in time-domain data with the lowest scores of both $MSE_{MAT}$ and $MSE_{TPSF}$ compared with two FEM simulators. Interestingly, we observed that the accuracy of calculated photon arrival time can be influenced more for different levels of light scattering. In contrast, the overall TPSF accuracy is more dependent on the scattering coefficient for both FEM simulators, which is similar to the observation in the SD precision analysis.

## 4. Discussion

This study systematically evaluates the performance of three mainstream open-source forward simulators (MMC, NIRFASTer, and Toast++) using experimental data acquired by a homebuilt TD-DOI system equipped with a SPAD array, with a particular focus on the impact of incident angle in a non-contact configuration. Prior validation efforts for SPAD-based TD systems have been limited in scope and often restricted to partial or single-metric evaluations (Jiang et al., 2021, Kalyanov et al., 2018). Thus, a quantitative, multi-metric, and cross-simulator assessment has been lacking. To address this gap, all effective TPSFs were considered, and 5 metrics were compared. We performed standardized homogeneous phantom experiments with unified mesh and source configurations, collecting and analyzing simulation and experimental data under two different incident angle conditions. A source correction strategy was adopted for FEM simulators, and the accuracy and applicability of each modelling approach were quantitatively assessed. The overall performance of the simulators is summarized across five key dimensions in figure 1b.

In terms of simulation settings, mesh resolution and time step size have a critical impact on NIRFASTer and Toast++. Coarse spatial meshes and large temporal steps can destabilize the TPSF profiles, leading to inaccuracies in temporal responses and ultimately increasing overall simulation error. Figure 2f shows that finer meshes reduce FEM simulation error by minimizing discretization artifacts, thereby yielding results closer to the underlying physical model. However, when the mesh is further refined, the error instead increases. A possible explanation is that the Gaussian-weighted computation of detected intensity may cause the improvement in mesh resolution to amplify local numerical fluctuations, thereby increasing the MSE. Moreover, Toast++ exhibits a larger decrease in TPSF MSE compared with NIRFASTer, which can be explained by the fact that NIRFASTer incorporates specific optimizations for TPSF in the early time regime, whereas Toast++ still shows noticeable fluctuations. Despite the difference, both FEM results, based on the diffusion equation with Robin boundary conditions, converged toward MC results, confirming their accuracy and reliability. It is also noteworthy that as the mesh density or the number of sources increases, the time cost of NIRFASTer, initially longer than that of Toast++, becomes shorter. This behaviour arises from the distinct preconditioning strategies employed by the two simulators: Toast++ typically relies on incomplete factorization methods, whereas NIRFASTer adopts the factored sparse approximate inverse (FSAI), which is faster and more amenable to parallelization. Furthermore, MMC requires a sufficient number of photons to ensure statistical convergence. Experimental results further demonstrate that the incident angle has a notable influence on the detected signals, which is an aspect often neglected in most forward modelling studies. By introducing a virtual source correction strategy, we verify that incorporating angle information under large-angle incidence can effectively enhance simulation accuracy for FEM-based simulations.

Regarding the accuracy of forward modelling, MMC exhibits the most accurate and robust performance across various error metrics, aligning well with its principle of directly tracing photon propagation paths to reflect the real physical process of RTE. In contrast, FEM simulators relying on numerical approximations of DE exhibit larger errors. The simulation error of NIRFASTer is close to that of Toast++ yet remains slightly lower. For both SD and TD precision metrics, FEM simulators tend to exhibit higher errors under low-scattering conditions. In the case of group A ($\mu_s' \approx 0.5$ mm$^{-1}$), the scattering coefficient may be too low to satisfy this prerequisite of DE. Under higher $\mu_s'$, its performance becomes comparable to others. These findings provide quantitative guidance for selecting appropriate forward models in practical applications. Analysis of TPSF curves at specific pixels reveals that, when FWHM is large, the simulated number of late-arriving photons tends to be lower than measured values (figure 5c). This discrepancy may be attributed to deviations between the assumed optical parameters and the actual optical properties of the tissue, leading to inaccuracies in modeling photon paths over longer distances.

The simulators assessed in our study are extensively employed in image reconstruction or evaluation tasks for various DOI modalities. Cooper et al utilized MMC to simulate photon propagation in head tissues and thereby compared the localization errors of TD-DOT under different guidance modes (Cooper et al., 2012). Jiang et al employed NIRFASTer to estimate optical parameters in vivo using TD-DOT (Jiang et al., 2021). Frijia et al used Toast++ to model optical fluence and investigate hemodynamic responses in infant brain imaging with HD-DOT (Frijia et al., 2021). As the accuracy of the simulators is intrinsically linked to the reliability of the resulting reconstructions, the present study fills the gap of comprehensive simulation validation in TD-DOI and enhances the confidence and accuracy of its future neuroimaging applications.

Despite the comprehensive analysis results, several limitations exist in this work. First, our work does not address complex heterogeneous structures or real biological tissues. Second, the proposed source correction method is based on an idealized model and needs to be validated in the context of image reconstruction. Moreover, the current correction model is limited to homogeneous planar geometries and requires generalization for anatomically complex biological structures. In future work, the virtual source correction will be extended to curved surfaces by incorporating surface-normal vector mapping. Lastly, the simulations were performed using a single point source, so future studies should consider implementing structured light (Lou et al., 2025) to enable faster data acquisition.

## 5. Conclusion

This study presents a comprehensive experimental validation of three time-domain diffuse optical simulators (MMC, NIRFASTer, and Toast++) using a SPAD-based TD-DOI system and standardized phantoms. MMC achieves the highest spatial-temporal accuracy, attributable to MC-based photon transport modelling, but is computationally expensive for multi-source cases. For FEM solvers, virtual source correction is crucial for oblique illumination ($\theta > 10°$), reducing SD/TD errors by > 34% in large-angle ($\theta = 58°$) scenarios. NIRFASTer and Toast++ demonstrate comparable overall performance, although NIRFASTer achieves slightly lower errors, particularly in the low-scattering regime. Overall, forward model selection should balance accuracy, efficiency, and scalability, tailored to the application, system, and computational resources.

**Acknowledgements**

The work is supported by Science and Technology Commission of Shanghai Municipality under Grant 25JS2830300 (Wuwei Ren), National Natural Science Foundation of China under Grants 62105205 (Wuwei Ren), and Shanghai Clinical Research and Trial Center (Wuwei Ren).

**Conflict of interest**

The authors declare no conflicts of interest. The authors would like to disclose that Edoardo Charbon is co-founder of PiImaging Technology and NovoViz. Both companies have not been involved with the paper drafting, and at the time of the writing have no commercial interests related to this article.